\documentclass[12pt,twocolumn]{emulateapj}

\shorttitle{Electron/Positron Excess from Pulsars}
\shortauthors{Kawanaka, Ioka and Nojiri}

\begin{document}

\title{Cosmic-Ray Electron Excess from Pulsars is Spiky or Smooth?: Continuous and
Multiple Electron/Positron injections}
\author{Norita Kawanaka\altaffilmark{1}, Kunihito Ioka\altaffilmark{1} and Mihoko M. Nojiri\altaffilmark{1,2}}
\altaffiltext{1}{Theory Center,
Institute of Particle and Nuclear Studies,
KEK (High Energy Accelerator Research Organization),
1-1 Oho, Tsukuba 305-0801, Japan} 
\altaffiltext{2}{Institute for the Physics and Mathematics of the Universe,
The University of Tokyo, Chiba 277-8568, Japan}
\email{norita.kawanaka@kek.jp}

\begin{abstract}
We investigate the observed spectrum of cosmic-ray electrons and positrons
from astrophysical sources, especially pulsars,
and the physical processes for making the spectrum spiky or smooth
via continuous and multiple electron/positron injections.
We find that (1) the average electron spectrum predicted from nearby pulsars
 are consistent with PAMELA, Fermi and H.E.S.S. data.  However,
the ATIC/PPB-BETS peak around 500GeV is hard to produce by the sum of
multiple pulsar contributions and requires a single (or a few)
energetic pulsar(s).
(2) A continuous injection produces a broad peak and
a high energy tail above the peak,
 which can constrain the source duration ($\lesssim 10^5$yr with the
current data).
(3) The H.E.S.S. data in the TeV range suggest that
young sources with age less than $\sim 6 \times 10^4$yr are
less energetic than $\sim 10^{48}{\rm erg}$.
(4) We also expect a large dispersion in the TeV spectrum
due to the small number of sources, that may cause the high
energy cutoff inferred by H.E.S.S. and potentially provide a
smoking-gun for the astrophysical origin.
These spectral diagnostics can be refined in the near future
by the CALET experiments to discriminate
different astrophysical and dark matter origins.
\end{abstract}

\section{Introduction}

Recently, the cosmic-ray positron fraction (the ratio of positrons to electrons plus positrons) has been measured by PAMELA satellite (Adriani et al. 2008).
  The observed positron fraction rises in the energy range of $10{\rm GeV}\lesssim \varepsilon_{e^{\pm}} \lesssim 100{\rm GeV}$, contrary to
 the prediction of secondary positrons, which are generated from cosmic rays propagating in the interstellar medium (ISM).  The ATIC
 balloon experiment has also revealed that there is an excess above $300{\rm GeV}$ and a possible peak at $\varepsilon_{e^{\pm}}\sim 600{\rm GeV}$
 (Chang et al. 2008), which is also reported by PPB-BETS (Torii et al. 2008b).  These observations
 strongly indicate nearby sources of $e^{\pm}$ pairs within $d\sim 1{\rm kpc}$ since high energy electrons/positrons lose their energy during propagation.
Possible candidates include a pulsar (Shen 1970; Chi et al. 1996; Zhang \& Cheng 2001; Grimani 2007; Kobayashi et al. 2004; B\"uesching et al. 2008; Hooper et al. 2009;
 Yuksel et al. 2008; Profumo 2008; Malyshev et al. 2009; Grasso et al. 2009), a microquasar (Heinz \& Sunyaev 2002),
 a gamma-ray burst (GRB; Ioka 2008), a supernova remnant (SNR; Shen \& Berkley 1968; Cowsik \& Lee 1979; Erlykin \& Wolfendale 2002; Pohl \& Esposito 1998;
 Kobayashi et al. 2004; Shaviv et al. 2009; Fujita et al. 2009; Hu et
  al. 2009; Blasi 2009; Blasi \& Serpico 2009; Mertsch \& Sarkar 2009;
  Biermann et al. 2009)
 and dark matter annihilations/decays (Asano et al. 2007; Arkani-Hamed et al. 2009;
 Bergstrom et al. 2008; Hamaguchi et al. 2008; Cirelli \& Strumia 2008; Cholis et al. 2008a, 2008b; Chen  et al. 2008, 2009a, 2009b; Chen \& Takahashi 2008;
 Hisano et al. 2005, 2008a, 2008b, 2009; Ishiwata et al. 2008a, 2008b,
  2008c; Zhang et al. 2008; March-Russell \& West 2008; Hooper et
  al. 2008; Pohl 2009).  Instead we might be observing the propagation
  effects (Delahaye et al. 2008; Cowsik \& Burch 2009; Stawarz et al. 2009) or the proton
  contamination (Fazely et al. 2009; Schubnell 2009).

  In order to discriminate different models of sources, an important diagnostic should be the spectral shape, in particular whether the ATIC/PPB-BETS peak
 is spiky or smooth.  In this regard, it is remarkable that an astrophysical source can make a peak with a sharp cutoff that is similar to the dark matter
 predictions, if the source is a transient object like a GRB (Ioka
 2008).  However, other astrophysical sources like pulsars, SNRs or microquasars are not transient
 and expected to have a finite spread in the cutoff, as suggested by Ioka (2008).  More importantly, due to the collimated emission, there are many off-axis
 pulsars that have not been observed via electromagnetic radiation, and we expect integrated contributions from multiple sources to the spectral shape, 
considering the birth rate of pulsars in our Galaxy.

In addition, recently the Fermi Large Area Telescope has measured the electron spectrum
 up to $\sim 1{\rm TeV}$ that is roughly proportional to $\sim \varepsilon_e^{-3}$ without any spectral peak as reported by ATIC/PPB-BETS (Abdo et al. 2009).
  The H.E.S.S. collaboration also provides the electron spectrum
 (Aharonian et al. 2008b, 2009), which is consistent with the Fermi
 result up to $\sim 1{\rm TeV}$ and shows the steep drop of the flux
 above that energy.  The Fermi data, however, should has a large
 systematic error in the high energy range ($\gtrsim 300{\rm GeV}$)
 where a significant
 fraction of electrons are removed to avoid a large hadron
 contamination, and so the real flux is estimated not by the pure
 experimental data but by the Monte Carlo
 simulations (Moiseev et al. 2007).  On the other hand the ATIC data
 contains the larger statistical errors than the Fermi data.  Therefore we
 cannot judge which observations are more reliable so far.

In this paper we investigate the effects of continuous and multiple pair injections on the observed electron/positron spectrum.
  We show that the flux above the peak energy does not drop off abruptly but remain finite if the pair injection continues for a finite time
 and suggest that we may measure the source duration from the peak
 width.  We also show that an average spectrum from multiple sources
 is relatively flat as reported by Fermi
 and the ATIC/PPB-BETS peak requires a single (or a few) extraordinary energetic source(s).
  We discuss the range of physical parameters of the sources
 (total electron/positron energy, the source duration, etc.) that are consistent with the current observational data.
 
\section{Injection Models and Calculations}
\subsection{Continuous $e^{\pm}$ Injection from a Single Source}
We assume that a point-like source starts injecting $e^{\pm}$ pairs at the time $t=0$ with total energy $E_{e^+}\sim E_{e^-}$ at a distance
 $d$ ($\sim 1{\rm kpc}$) from the Earth.  The observed electron/positron spectrum after the propagation is obtained by solving the diffusion equation,
\begin{eqnarray}
\frac{\partial}{\partial t}f=K(\varepsilon _e)\nabla ^2 f+\frac{\partial}{\partial \varepsilon _e}[B(\varepsilon _e)f]+Q(t, \mbox{\boldmath $r$}, \varepsilon _e), \label{diff}
\end{eqnarray}
where $f(t, \mbox{\boldmath $r$}, \varepsilon _e)$ is the distribution function of particles at time $t$ and position $\mbox{\boldmath $r$}$ with energy $\varepsilon _e$.
  Here $K(\varepsilon_e)=K_0 (1+\varepsilon _e/3{\rm GeV})^\delta$ is the diffusion coefficient, $B(\varepsilon _e)$ is the energy loss rate, and $Q$ is
 the injection rate of electrons/positrons.
  Hereafter we adopt $K_0 =5.8 \times 10^{28} {\rm cm^2~s^{-1}}$, $\delta=1/3$ that is consistent with the boron/carbon ratio according to the latest GALPROP code, and
 $B(\varepsilon_e)=-b\varepsilon_e^2$ with $b=10^{-16}{\rm GeV^{-1}~s^{-1}}$ which includes the energy loss due to synchrotron emission and inverse Compton scattering
 (Baltz \& Edsj\"o 1999; Moskalenko \& Strong 1998).

Here we assume the continuous injection with a power-law spectrum:
 $Q(t, \mbox{\boldmath $r$}, \varepsilon _e)\propto Q_0(t)\varepsilon_e^{-\alpha} \delta(\mbox{\boldmath $r$}-\mbox{\boldmath $r$}_0)$.
   We can obtain the observed spectrum for an
 arbitrary type of injection using the Green's function of the equation (\ref{diff}), derived in Atoyan et al. (1995), with respect to $\mbox{\boldmath $r$}$ and $t$:
\begin{eqnarray}
G(t, \mbox{\boldmath $r$}, \varepsilon_e;t_0, \mbox{\boldmath $r$}_0)
&=&\frac{Q_0(t_0)\varepsilon_{e,0}^{-\alpha}B(\varepsilon_{e,0})}{\pi^{3/2}B(\varepsilon_e)d_{\rm diff}^3} \nonumber \\
&&\times \exp \left( -\frac{r^2}{d_{\rm diff}^2}\right), \label{green}
\end{eqnarray}
where $\varepsilon_{e,0}=\varepsilon_e/[1-b(t-t_0)\varepsilon_e]$ is the energy
 of electrons/positrons at the time $t_0$ which are cooled down to $\varepsilon_e$ at the time $t$, $r=|\mbox{\boldmath $r$}-\mbox{\boldmath $r$}_0|$, and $G=0$ when $\varepsilon_{e,0}$
 is larger than the maximum energy of
 the injection spectrum, $\varepsilon_{e,{\rm max}}$.  As $B(\varepsilon_{e,0})/B(\varepsilon_e)=(\varepsilon_{e,0}/\varepsilon_e)^2<(\varepsilon_{e,{\rm max}}/\varepsilon_e)^2<\infty$,
 there is no divergence in Eq.~(\ref{green}).  We can approximate the diffusion length as
\begin{eqnarray}
d_{\rm diff}\simeq 2\sqrt{K(\varepsilon_e)(t-t_0)\frac{1-(1-\varepsilon/\varepsilon_{\rm cut})^{1-\delta}}{(1-\delta)\varepsilon_e/\varepsilon_{\rm cut}}}.
\end{eqnarray}
 when $\varepsilon_e \gg 3{\rm GeV}$ and the diffusion coefficient is almost power-law $K(\varepsilon_e)\simeq K_0(\varepsilon_e/3{\rm GeV})^{\delta}$.  Here
 $\varepsilon_{\rm cut}=[b(t-t_0)]^{-1}$.

Once we assume the injection rate $Q_0(t_0)$, we can obtain the observed electron/positron
 spectrum by integrating (\ref{green}) over $t_0$:
\begin{eqnarray}
f(t, r, \varepsilon_e)=\int_{t_i}^{t}G(t,\mbox{\boldmath $r$},\varepsilon_e;\tau, \mbox{\boldmath $r$}_0)d\tau.
\end{eqnarray}

Note that the initial time of the integration should be set as $t_i={\rm max}[0, t-b^{-1}(\varepsilon_e^{-1}-\varepsilon_{e,{\rm max}}^{-1})]$.

We consider two types of continuous injection.  One is the pulsar-type decay:
\begin{eqnarray}
Q_0(\tau)\propto \frac{1}{(1+\tau/\tau_0)^2}. \label{pulsartype}
\end{eqnarray}

This is the similar function of time as the spin-down luminosity of a pulsar with a surface magnetic field
\begin{eqnarray}
B=8.6\times 10^{11} P_{10{\rm msec}} \left(\tau_{0,4}\right)^{-1/2},
\end{eqnarray}
where $P_{10{\rm msec}}$ is the pulsar period normalized by 10msec and $\tau_{0,4}=\tau_0/10^4{\rm yr}$ (Shapiro \& Teukolsky 1983).  The other is the exponential decay:
\begin{eqnarray}
Q_0(\tau)\propto \exp \left(-\frac{\tau \ln 4}{\tau_0}\right). \label{exptype}
\end{eqnarray}
which may be realized by a pulsar that initially confines $e^{\pm}$ in
 its nebula and releases them afterward, by a SNR that
 accelerates protons and continues to inject them to the surrounding dense gas cloud until
 it is destroyed, or by a microquasar ceasing its activity.  In
 both types of injection, the characteristic time scale of the duration $\tau_0$ is defined to be the time when the rate becomes four times smaller than
 the initial one.

In Fig.~1 we show the electron plus positron flux resulting from above two injection models in addition to the transient model ($\tau_0=0$)
and the background\footnote{For the background shown in the following plots, we adopt the fitting functions in Baltz \& Edsj\"o (1999) by reducing the primary $e^-$ flux,
 which is conventionally attributed to supernova remnants, by 30\% because the fitting functions provide larger flux than the ATIC data even without other contributions.} (dotted line).  
The remarkable point is that an astrophysical source can make a spectral peak that is similar to the ATIC/PPB-BETS
 excess and also to the dark matter case (Ioka 2008).  The peak energy is determined by the age of the source $t_{\rm age}$ as
\begin{eqnarray}
\varepsilon_{e, {\rm peak}}=\left[ bt_{\rm age}+\frac{1}{\varepsilon_{e,{\rm max}}}\right]^{-1}, \label{peak}
\end{eqnarray}
because the electrons/positrons with initially higher energy cool down
 via synchrotron emission and inverse Compton scattering within time $t_{\rm age}$.  We can inversely
 estimate the source age as $t_{\rm age}\sim 5\times 10^5{\rm years}$ from the peak energy for $\varepsilon_{e,{\rm max}}\gtrsim 1{\rm TeV}$.
  Note that the peak flux is almost independent of the distance $r$ if it is smaller than the diffusion length ($\sim 1{\rm kpc}$ in our case).

  As is clear from Fig.~1, the spectral cutoff becomes shallower for the continuous injection models than the transient one ($\tau_0=0$; short dot-dashed line).
  This is because the significant fraction of $e^{\pm}$ pairs are produced recently
 (i.e. injected long after the birth of the source) and they have
  shorter time for the energy loss via synchrotron emission and
  inverse Compton scattering.
  Then their energy is still higher than the peak energy when they reach the Earth, and they produce a broader peak.

\begin{figure}

\plotone{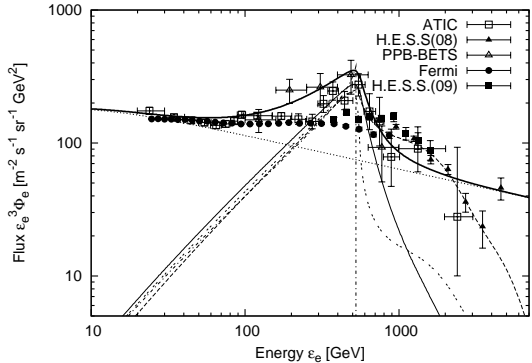}

\caption{The electron plus positron flux 
predicted from a source that continuously injects pairs for a finite duration $\tau_0=10^5{\rm years}$ with the exponential decay in Eq.~(7) (thin solid line),
and its sum (thick solid line) with the background (dotted line),
compared with the ATIC/PPB-BETS/H.E.S.S./Fermi data.
We also show the pulsar-type injection in Eq.~(5) with $\tau_0=10^5{\rm years}$
(long dashed line) and $\tau_0=10^4{\rm years}$ (double dashed line), 
in addition to the transient injection ($\tau_0=0$; short dot-dashed line).
We assume that a source at $r=1{\rm kpc}$
from the Earth a time $t_{\rm age}=5.6\times 10^5{\rm years}$ ago produces $e^{\pm}$ pairs with total energy $E_{e^+}=E_{e^-}=0.8\times 10^{50}{\rm erg}$
and spectral index $\alpha=1.7$ up to $\varepsilon_{e,{\rm max}}=10{\rm TeV}$.
}

\label{f1}

\end{figure}

The thick solid line represents the total (the primary plus background electron and positron) flux assuming that the source starts emitting
 $e^{\pm}$ pairs with total energy $\sim 10^{50}{\rm erg}$, a power-law index $\alpha\sim 1.7$ and a maximum energy $\sim 5{\rm TeV}$ at a distance
 $\sim 1{\rm kpc}$ from the Earth a time $t_{\rm age}\sim 5\times 10^5 {\rm yr}$ ago, and decays exponentially with the duration of
 $\tau_0\sim 10^5{\rm year}$.  This model looks better for the ATIC/PPB-BETS peak, though we cannot
 conclude that the duration is finite with the current data.  The positron fraction predicted from this parameter set is also consistent
 with the PAMELA results, in almost the same way as Fig.~1 of Ioka (2008).

In the case of pulsar-type injection, there is another interesting spectral feature resulting from a long duration.  In Fig.~1, the high energy tail
 above the peak energy is more enhanced for the long duration case ($\tau_0=10^5{\rm years}$, double dashed line) than the short duration case
 ($\tau_0=10^4{\rm years}$, long dashed line).  This is because the longer the duration
 of injection is, the larger fraction of $e^{\pm}$ pairs are freshly produced and they do not lose their energy during the propagation so much
 (see also Atoyan et al. 1995).  Especially, the flux of the long duration model may exceed the H.E.S.S. observations around $\sim 4{\rm TeV}$
 if we add the background (dotted line) while that of the short duration model does not.  As the errorbars are still large, however, we should
 await future observations.

\subsection{Multiple $e^{\pm}$ Injections: Average Flux and Its Dispersion}
Next, let us consider multiple sources.  We expect several younger or older pulsars than that in Fig.~1 of age $t_{\rm age}\sim 5\times 10^5{\rm yr}$,
 considering the local birth rate of pulsars $\sim 10^{-5}{\rm
 yr}^{-1}{\rm kpc}^{-2}$ (Narayan 1984; Lorimer et al. 1993).
  Moreover, the total energy $E_{e^+}+E_{e^-}$ in Fig.~1 is as large as the rotation energy
 of a pulsar $E_{\rm rot}$ with a period of $\sim 10{\rm msec}$.  This is comparable with the fastest initial spin estimated from the observations of radio pulsars
 (Kaspi \& Helfand 2002), so the pair output efficiency $f_e\equiv
 (E_{e^+}+E_{e^-})/E_{\rm rot}$ may be too large $\sim 100\%$ to account for the excess with a single pulsar. 

We can calculate the average electron and positron spectrum by considering the nearby multiple pulsars with a certain birth rate in the following way.  Once $\varepsilon_e$ is fixed,
 we can neglect the contribution from pulsars older than $\sim 1/(b\varepsilon_e)$ and farther than $\sim d_{\rm diff}\sim 2\sqrt{K(\varepsilon_e)t}$
 as is obvious from the functional form of Eq.(\ref{green}).  Then
 the average flux can be calculated by:
\begin{eqnarray}
f_{\rm ave}(\varepsilon_e)&=&\int_0^{1/(b\varepsilon_e)}dt\int_0^{d_{\rm diff}}2\pi r dr f(t,r,\varepsilon)R \label{average} \nonumber \\
&\sim&\frac{Q_0 R}{\pi^{1/2}\sqrt{K(\varepsilon_e)b\varepsilon_e}}\varepsilon_e^{-\alpha}, \nonumber \\
&=&N(\varepsilon_e)\times f_{\rm 1,ave}(\varepsilon_e), \label{ave}
\end{eqnarray}
where $R$ is the local pulsar birth rate (${\rm yr}^{-1}{\rm
kpc}^{-2}$),
\begin{eqnarray}
N(\varepsilon_e)&=&\int_0^{1/(b\varepsilon_e)}dt\int_0^{d_{\rm
 diff}}dr2\pi rR \nonumber \\
&\sim&\frac{2\pi K(\varepsilon_e)R}{\left( b\varepsilon_e \right) ^2},
\end{eqnarray}
is the number of pulsars which contribute to the flux at the energy
$\varepsilon_e$, and $f_{\rm 1,ave}(\varepsilon_e)=f_{\rm
ave}(\varepsilon_e)/N(\varepsilon_e)$ is the average electron flux per
pulsar.  Here we adopt the value of $R$ as the birth rate per
unit surface area because pulsars are born from a disk whose
thickness ($\sim 200-300{\rm pc}$) is much smaller than the diffusion
length of cosmic-rays ($\sim 2-3{\rm kpc}$). 

Fig.~2 and Fig.~3 show the average electron spectra and the positron
fraction, respectively, obtained by assuming that each pulsar emits
electrons with the total amount of energy of $\sim 1\times 10^{48}{\rm erg}$, the spectral index of $\sim
1.9$, and the birth rate of $R\sim 1/(1.5\times 10^5) {\rm
yr}^{-1}{\rm kpc}^{-2}$ (thick solid lines).

In addition, we can calculate the dispersion of the number of pulsars $\Delta N(\varepsilon_e)$ from the average $N(\varepsilon_e)$
 for each energy bin as
\begin{eqnarray}
\Delta N(\varepsilon_e)\sim \sqrt{N(\varepsilon_e)},
\end{eqnarray}
which is based on the Poisson distribution of nearby pulsars.  Then we
can estimate the flux dispersion as
\begin{eqnarray}
\Delta f_{\rm ave}(\varepsilon_e)\sim f_{\rm
1,ave}(\varepsilon_e)\sqrt{N(\varepsilon_e)}=f_{\rm
ave}(\varepsilon_e)/\sqrt{N(\varepsilon_e)}. \label{numdisp}
\end{eqnarray}

From Fig.2 and 3 we can see that the average spectra are basically consistent with Fermi, H.E.S.S. and PAMELA data.  In the high energy range ($\varepsilon_e\gtrsim {\rm TeV}$),
 the dispersion from the average flux become significant.  This can be interpreted as follows.  The pulsars which contribute to the electron and positron
 flux in such a high energy band should be young,
\begin{eqnarray}
t_{\rm age}\lesssim \frac{1}{b\varepsilon_e}\sim 3.1\times 10^5{\rm yrs}\left( \frac{\varepsilon_e}{\rm TeV}\right)^{-1},
\end{eqnarray}
and close to the Earth,
\begin{eqnarray}
r\lesssim 2\sqrt{K(\varepsilon_e)t_{\rm age}}\sim  1.3{\rm kpc}\left( \frac{\varepsilon_e}{\rm TeV}\right)^{-1/3},
\end{eqnarray}
where we adopt $\delta=1/3$.  The number of such pulsars should be as small as
\begin{eqnarray}
N(\varepsilon_e)\sim 6 \left(\frac{\varepsilon_e}{\rm
TeV}\right)^{-5/3}\left( \frac{R}{1/(1.5\times 10^5){\rm yr}^{-1}{\rm kpc}^{-2}}\right).\label{number}
\end{eqnarray}

Therefore, in the TeV range few pulsars can contribute to the
 electron/positron flux.  This small number of pulsars may
 naturally account for the spectral cutoff around $\sim {\rm
 TeV}$ energy, which has been inferred by the H.E.S.S.
 observations.  Strictly speaking, this estimation of the pulsar number
 dispersion is not correct in the energy range of $\varepsilon_e
 \gtrsim 3{\rm TeV}$, where $N(\varepsilon_e) \lesssim 1$ and the
 statistical arguments become meaningless.  However this interpretation
 of the spectral drop around this energy is still qualitatively correct.

Moreover, Fig.2 shows that the
  ATIC/PPB-BETS peak flux ($\varepsilon_e\sim 545{\rm GeV}$) is much
  larger than the average flux added with the dispersion flux $\Delta
  f_{\rm ave}$ at the same energy bin.  In fact, the separation
  between the average flux and the ATIC data of the peak flux at that
  energy is $\sim 10\Delta f_{\rm ave}$.  Then, if all pulsars emit
  electrons with the total energy of $\sim 10^{48}{\rm erg}$, the
  number of pulsars which contribute to the energy bin of the
  ATIC/PPB-BETS should be unrealistically large at the 10$\sigma$ level.  This means that if the
 ATIC/PPB-BETS peak is real, it does not seem to be produced by the collective
 contribution from multiple pulsars with the moderate amount of electron energy ($\sim
 10^{48}{\rm erg}$) but by a single (or a few) energetic
  pulsar(s) ($\sim 10^{49-50}{\rm erg}$)\footnote{Since the multiple
  contributions tend to make the spectrum softer, it is possible to fit the ATIC/PPB-BETS spectrum with multiple pulsars by using the harder
  spectral index $\alpha$ accordingly.  However, in order to fit the PAMELA spectrum in the lower energy range at the same time, such a hard spectrum is not favored.}.  

The discussion above is about the fluctuation of the number of pulsars with a certain birth rate, and it is based on the Poisson statistics.
  Strictly speaking, in order to discuss the cosmic ray
  electron/positron fluctuations due to the random injections, one should evaluate not the dispersion of the source number
 but the dispersion of the electron/positron flux at each energy bin as
\begin{eqnarray}
\Delta f_{\rm ave}^2 \sim \int_0^{1/(b\varepsilon_e)}dt \int_0^{d_{\rm diff}}2\pi rdr f^2 R-N(\varepsilon_e)f_{\rm 1, ave}^2. \label{fluxdisp}
\end{eqnarray}

The first integral in Eq.(\ref{fluxdisp}) contains, however, a serious divergence because of the large (but improbable) contribution from very young and nearby sources
 (Lee 1979; Berezinskii et al. 1990; Lagutin and Nikulin 1995; Ptuskin et al. 2006).
  In order to obtain the realistic estimate of the flux dispersion, we introduce a lower cutoff parameter $\tau_c$ to the time integral.  Then we have
\begin{eqnarray}
\Delta f_{\rm ave}^2 \sim \frac{Q_0^2 R}{16\pi^2 K(\varepsilon_e)^2\tau_c}\varepsilon_e^{-2\alpha}.
\end{eqnarray}

  Following Ptuskin et al. (2006), we adopt the cutoff parameter as
\begin{eqnarray}
\tau_c&=&[4\pi R K(\varepsilon_e)]^{-1/2} \nonumber \\
&\simeq& 10^5{\rm yr}\left(\frac{\varepsilon_e}{500{\rm
GeV}}\right)^{-1/6}\left( \frac{R}{1/(1.5\times 10^5){\rm yr}^{-1}{\rm
kpc}^{-2}} \right)^{-1/2},
\end{eqnarray}
which takes into account the absence of very young and nearby sources.
  This choice of $\tau$ is reasonable as long as this time is much
  shorter than $(b\varepsilon_e)^{-1}$, which means $\varepsilon_e \ll 4.4{\rm TeV}$.
  Then the ratio of the flux dispersion to the average flux
  (\ref{ave}) can be expressed as
\begin{eqnarray}
\frac{\Delta f_{\rm ave}}{f_{\rm ave}}&\sim& 0.21
  \left(\frac{\varepsilon_e}{500{\rm GeV}}\right)^{5/12} \nonumber \\
&\times& \left(\frac{R}{1/(1.5\times 10^5){\rm yr}^{-1}{\rm kpc}^{-2}}\right)^{-1/4}.\label{fluxdisp2}
\end{eqnarray}

  We can see that these two `dispersions' (Eq.(\ref{numdisp})
  and Eq.(\ref{fluxdisp2})) give the similar results
  around the energy of the ATIC/PPB-BETS peak.  Therefore, from either
  of the above discussions, we can say that the ATIC/PPB-BETS data of the spectral peak
 is so largely separated from the average flux that they do not seem to be produced by the multiple contribution from nearby pulsars with moderate energy.



  We should note that the H.E.S.S. data put constraints on the total $e^{\pm}$ pair energy from young sources.
  We plot in Fig.~2 the electron spectrum from the source with the age of $\sim
  6\times 10^4{\rm years}$ so as not to exceed the observational
  upper limit inferred by the H.E.S.S. data in the TeV range
  (long-dashed line).  We find that the total energy of such young
  sources should be, if exist, $\lesssim 2\times 10^{48}{\rm erg}$
  which is two orders-of-magnitude smaller than the energy of the
  source making the ATIC/PPB-BETS peak.

   
\begin{figure}

\plotone{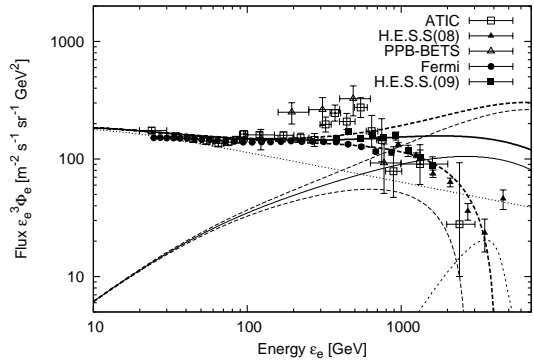}

\caption{The average electron plus positron flux (thick solid line),
  the flux with the standard deviation ($f_{\rm ave}\pm \Delta f_{\rm
 ave}$, upper/lower thick dashed lines) with the background (dotted line) predicted
 from the local pulsar birth rate of $\sim 1/(1.5\times 10^5)$ per year
 per ${\rm kpc}^2$ and the total electron energy of $\sim 10^{48}{\rm erg}$ (thin solid line and upper/lower thin dashed lines for
 the same set of the spectra without the background), compared with the ATIC/PPB-BETS/H.E.S.S./Fermi data.
  The flux from a young source ($\sim 6\times 10^4{\rm years}$) with the energy of $\sim 2\times 10^{48}{\rm erg}$ (long-dashed line) is also shown.
  We assume that each source emits
 electrons/positrons with a power-law index $\alpha=1.9$ up to $\varepsilon_{e,{\rm max}}=10{\rm TeV}$.}

\label{f2}

\end{figure}

\begin{figure}

\plotone{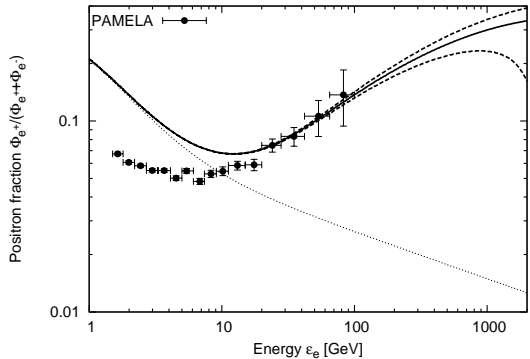}

\caption{The total positron fraction resulting from the average spectrum (thick solid line) and the dispersion (thick dotted lines), which have the
 same parameters as in Fig.~2, and the background (dotted line), compared with the PAMELA data.  Note that the solar modulation is important below $\sim 10{\rm GeV}$.}

\label{f3}

\end{figure}

\section{Discussion and Conclusion}

We investigate the astrophysical origin for the PAMELA and ATIC/PPB-BETS excesses and in particular the effects of the finite duration and the multiple
 sources on the electron and positron spectra, as expected for
 pulsars, SNRs and microquasars.  We find the followings:
 
(1) A non-transient source can make a spectral peak that is similar to the ATIC/PPB-BETS excess (see Fig.~1) around the peak energy in Eq.~(\ref{peak}).
  The peak is generally broad with a width
\begin{eqnarray} 
\left| \frac{\Delta \varepsilon_{e, {\rm peak}}}{\varepsilon_{e, {\rm peak}}}\right| \approx \frac{\tau_0}{t_{\rm age}}
\sim 10\%~{\tau_{0,4}}\left(\frac{t_{\rm age}}{10^5{\rm years}}\right)^{-1},
\end{eqnarray}
which could provide a method to measure the source duration $\tau_0$ by the Fermi satellite (an energy resolution of 5-20\% in 20GeV-1TeV range; Moiseev et al. 2007)
 or the future CALET experiments (a few \% above 100GeV; Torii et
 al. 2008a).  Although Atoyan et al. (1995) have already pointed out
 the effects of finite duration of the source on the electron
 spectrum, they only mention the enhancement of the high energy tail
  above the spectral peak (see below) and never discuss the peak
 width.  Note that the peak width is also produced by the spatial fluctuation of Galactic
 magnetic field and the photon density because the energy loss rate of $e^{\pm}$ fluctuates during the propagation, as estimated in Ioka (2008) (see also Malyshev et al. 2009).  We also note that the peak becomes smoother
if the injection rises gradually in its initial stage.

(2) The spectrum from a long duration source has a high energy tail
  above the peak energy (see Fig.~1).  Especially the flux of this tail plus the background may exceed the H.E.S.S. data points  when assuming a pulsar-type decay with a duration $\tau_0\gtrsim 10^5{\rm years}$.  This implies that 
 the source making the ATIC/PPB-BETS peak is not likely a single pulsar with magnetic fields weaker
  than a few times $10^{11}{\rm G}$.  The existence of
  this tail has been already pointed out before (Atoyan et al. 1995).
  However, we firstly present the quantitative
  argument for the observational limit of the duration of the
  electron/positron source in the context of the high energy tail thanks to the observational developments in the TeV range.  One should note that we cannot rule out the long-duration pulsar model if the maximum energy of injected $e^{\pm}$ pairs is smaller than
 $\lesssim {\rm TeV}$, or the injection is not the pulsar-type in
  Eq.~(5) but the exponential-type in Eq.~(7), for example.  The latter is possible
 if high energy pairs generated in
 the pulsar magnetosphere are not injected into the space instantaneously but initially confined in a pulsar wind nebula (Chi et al. 1996) and they diffuse out after the nebula gets broken.

(3) The H.E.S.S. data suggest that young sources with age less than
 $6\times 10^4{\rm yr}$ should be, if exist, two
 orders-of-magnitude less energetic than the source making the
 ATIC/PPB-BETS peak.  Note
 that the lifetime of the pulsar nebula is around $\sim 10^5{\rm yr}$
 and younger pulsars may not be able to contribute by the cosmic-ray
 confinement in the nebula.

(4) The average electron spectrum and positron fraction is well
  consistent with the H.E.S.S./Fermi and PAMELA data, respectively, taking
 into account the dispersion predicted from the total electron energy per pulsar of $\sim 10^{48}{\rm erg}$ with the local birth
 rate of $\sim 1/10^5{\rm yrs}/{\rm kpc}^2$.
  Especially, when $\varepsilon_e\gtrsim {\rm TeV}$, we expect a large dispersion of the electron flux
 because of the small number of sources which are young and close to the Earth and can significantly contribute to that energy range.  This fact can
 naturally account for the spectral drop around $\gtrsim {\rm TeV}$ indicated by the H.E.S.S. observations.  Note that the value of the total electron energy
 per pulsar adopted here is within reasonable range.  In fact,
 the pulsar whose initial spin period is around $\sim 10{\rm msec}$
  can emit electrons and positrons with energy $\sim 10^{48}{\rm
  erg}$ if we assume the
  efficiency of $f_e\sim 1\%$, which seems
 to be reasonable (Hooper et al. 2009).
 
  Moreover, we show that the ATIC/PPB-BETS data point
 showing the peak at $\varepsilon_e\sim 545{\rm GeV}$ is largely
 separated from the average flux, when considering the theoretical
 dispersion from the average.  This fact suggests that the peak
 is hard to produce by multiple contributions and requires a single (or a few) extraordinary
 pulsar(s) whose total electron/positron energy is about a hundred (several tens) times larger than that of ordinary pulsars.
  Here we estimate the dispersion of the electron/positron flux based
 on the analytical expressions (Eq. (9),(12) or (17)) using the averaged local birth rate of pulsars.
  This method enables us to take into account the off-axis pulsars whose existence is suggested by the observed pulse shape of pulsars,
 and it is different from the method used in Malyshev et al. (2009) who calculate some
 realizations of the spectra predicted from the known pulsars in the
 ATNF catalogue.


 
  Note that the different choice of
 the diffusion coefficient $K(\varepsilon_e)$ would change the results quantitatively.  The smaller $K$ makes the diffusion length $r_{\rm diff}$ smaller, and
 the particle density inside that radius gets higher, being proportional to $r_{\rm diff}^{-3}$ (see Eqs.(2) and (3)). For different $\tilde{K}$
 instead of $K$, we can apply our results by re-scaling the distance of each pulsar and the total $e^{\pm}$ injection energy as $d\rightarrow d\sqrt{\tilde{K}/K}$ and
 $E_{\rm tot}\rightarrow E_{\rm tot}(\tilde{K}/K)^{3/2}$, respectively.

In our calculations we evaluate the dispersion of the electron
 flux due to the random birth of nearby pulsars in time and space
 having uniform total energy and injection index.  In the case that
 these pulsars have a distribution of energy with a dispersion of
 $\delta E$, the total dispersion of the energy is averaged as $\sim
 \delta E\sqrt{N(\varepsilon)}$, and when the electron energy is smaller than $\sim {\rm TeV}$
 (i.e. $N(\varepsilon)$ is much larger than unity) the total dispersion is
 suppressed compared to the total flux $N(\varepsilon_e)f_{\rm 1,ave}$.
  The spectral index of the injected electrons should also be varied.
 However, the dispersion of the flux is almost determined by the
 amount of the electron energy emitted from pulsars, and the
 fluctuation of the index would not contribute to the flux dispersion so much.

  The spatial variation of the energy loss rate
 and the diffusion coefficient can also affect to the observed electron flux or positron fraction.
  The energy loss rate $b$ can fluctuate along the propagation path of electrons because of the inhomogeneities of the radiation and magnetic field,
 and then the cutoff shape of the resulting electron spectra would be broadened according to the amplitude of the fluctuation.
  Such a feature may be resolved by the future CALET experiment (see Ioka 2008).  On the other hand, the effects of the spatial variation
 of the diffusion coefficient are considered in Cowsik \& Burch (2009) in the context of "Nested leaky box model".  In this model the positron
 fraction can be explained as a result of the different diffusion coefficient between the source-surrounding region and the general interstellar space.

We can expect gamma-ray emission from high energy $e^{\pm}$ pairs.  Especially, the number of such energetic objects can be simply estimated as
 $\sim (10{\rm kpc}/1{\rm kpc})^2=100$.  This is comparable with that of TeV unidentified sources, which have no clear counterpart at other wavelengths
 (Aharonian et al. 2005, 2008a; Mukherjee and Halpern 2005; Ioka \& M\'esz\'aros 2009), implying some connections between them.

We thank H. Kodama and F. Takahashi for useful discussions.  We also acknowledge helpful comments and suggestions from an anonymous referee.
  This work is supported in part by the World Premier International Center Initiative
(WPI Program), MEXT, Japan and the Grant-in-Aid for Science Research, Japan Society
for the Promotion of Science (No. 18740147 and No. 19047004 for KI, No. 16081207 and No. 18340060 for MN).

\end{document}